# Attosecond pulse formation in the "water window" range by optically dressed hydrogen-like plasma-based $C^{5+}$ X-ray laser

V. A. Antonov[1,2,*], I. R. Khairulin[1,3], and Olga Kocharovskaya[4]

[1] *Institute of Applied Physics of the Russian Academy of Sciences,*
*46 Ulyanov Street, Nizhny Novgorod, 603950, Russia,*
[2] *Prokhorov General Physics Institute of the Russian Academy of Sciences,*
*38 Vavilov Street, Moscow, 119991, Russia,*
[3] *N.I. Lobachevsky State University of Nizhny Novgorod,*
*23 Gagarin Avenue, Nizhny Novgorod, 603950, Russia,*
[4] *Department of Physics and Astronomy and*
*Institute for Quantum Studies and Engineering,*
*Texas A&M University, College Station, TX 77843-4242, USA.*

In this paper, we present the analytical theory of attosecond pulse formation via optical modulation of an active medium of the hydrogen-like $C^{5+}$ plasma-based X-ray laser at 3.4 nm wavelength in the "water window" range, taking into account a variation of the population inversion caused by radiative decay of the upper lasing states. We derive an analytical solution for the X-ray field amplified by an X-ray laser with time-dependent population inversion, which is simultaneously irradiated by a strong optical laser field, and use it to find the optimal conditions for the attosecond pulse formation from a narrowband seeding X-ray field. We show that the shape of pulses can be improved at the cost of reduced pulse peak intensity (i) via external attenuation of the resonant spectral component of the amplified X-ray field or (ii) by using a resonantly absorbing medium (the active medium of the X-ray laser after the change of sign of the population inversion) for the pulse formation. The results of the analytical theory are in a good agreement with the numerical solutions of the Maxwell-Bloch equations which account for the nonlinearity, as well as the amplified spontaneous emission, of the active medium. Both analytically and numerically we show the possibility to produce a train of attosecond pulses with sub-200 as duration and the peak intensity exceeding $10^{12}$ W/cm$^2$ at the carrier wavelength 3.4 nm in the "water window" range, which makes them attractive for the biological and medical applications.

## I. INTRODUCTION

The time-resolved X-ray diffraction, absorption, and scattering are indispensable tools for study of the ultrafast electron dynamics and nuclear motions in atoms and molecules at their intrinsic time scales; as well as charge migration, structural and spin dynamics in solids; and photoinduced chemical reactions (see the recent reviews [1-6]). In particular, the time-resolved imaging of nanoscale biological materials and living cells attracts an increasing attention [1, 3-5]. In all of these applications a combination of the high temporal and spatial resolution with the high brightness of the X-ray field is desirable.

Up to date, the most promising (but at the same time the most demanding and expensive) sources of X-ray radiation with up to multi-keV photon energy are the X-ray free-electron lasers (XFELs) [4, 7], which are capable of producing the radiation of high brightness in a sequence of femto- or attosecond [8] bursts. But at the same time XFELs lack temporal coherence and reproducibility even in the self-seeded mode of operation, and there are only few such facilities in the world. The fully coherent X-ray pulses with photon energies exceeding 1 keV [9] and pulse duration down to 50 as [10] have been produced via the high-harmonic generation (HHG) of infrared (IR) laser fields in gases (see also [5] and [7]). However, although remaining the only reliable source of attosecond pulses, suitable for the practical applications, HHG is strictly limited in terms of pulse energy in the X-ray range (typically, by few or few tens of pJ for the photon energies exceeding 200 eV). The other widespread coherent X-ray sources are the plasma-based soft-X-ray lasers [11, 12], which allow producing the pulses of X-ray radiation with up to few μJ energy at a few nm wavelength [12-14] in a table-top setup. But the radiation bursts of the plasma-based X-ray lasers have ps or longer duration, which limits their applications for the time-resolved measurements.

Recently, we suggested an approach, which might allow to combine the merits of the HHG and the plasma-based X-ray sources, that is the attosecond pulse duration and the high radiation energy [15, 16]. The basic idea is to enrich the spectrum of generation and amplification of an X-ray laser by multiple equidistant sidebands via irradiation of its active medium by a strong optical laser field. The sidebands appear due to sub-optical-cycle modulation of the frequency of the inverted X-ray transition by the optical laser field via Stark splitting of the excited energy levels of the resonant ions [17, 18]. Under a proper choice of parameters of the modulating optical field and the active medium, the generated [15] or amplified [16] spectrum might be sufficiently broad and phase-aligned, so that the output X-ray field constitutes a train of attosecond pulses in the time domain. In the following work [19] we described the spectrum enrichment process analytically and found the optimal conditions for the transformation of a quasi-monochromatic seeding extreme ultraviolet (XUV) radiation into a sub-fs pulse train in active medium of a hydrogen-like $Li^{2+}$ soft-X-ray laser, dressed by an optical laser field (the resonant radiation wavelength in such a case is 13.5 nm) [20].

In the present contribution, we analyze the ultimate capabilities and limitations of this method in the X-ray range by considering the possibility to produce an attosecond pulse train from a quasi-monochromatic seeding X-ray field in active medium of the hydrogen-like $C^{5+}$ X-ray laser (the resonant radiation wavelength is 3.4 nm) [14]. In the case of $C^{5+}$ active medium (i) the lifetime of the population inversion is 16 times smaller than that in the case of $Li^{2+}$ plasma [19], while (ii) the gain for the resonant X-ray field is weaker, and (iii) the plasma dispersion for the modulating optical field is stronger. As a result, (a) it is necessary to account for the time-variation of the population inversion at the lasing transition, and (b) the sidebands are generated less efficiently than in the case of $Li^{2+}$ active medium. Therefore, formation of the high-contrast attosecond pulses implies either an attenuation of the resonant spectral component of the amplified X-ray field to the level of the generated sidebands, or using of a passive (absorbing), rather than active (amplifying) plasma medium. These questions are addressed in the present paper both analytically and numerically.

The paper is organized as follows. In Sec. II we briefly formulate the basic set of equations for the resonant X-ray field and the active medium (which is the same as in [19]). In Sec. III we introduce the analytical solution for the X-ray field (which is more general as compared to that in [19], as it takes into account the variation of population inversion). In Sec. IV based on both the analytical solution and numerical studies we find the optimal conditions for the formation of attosecond pulses with the highest contrast in either active or passive medium. Also, we discuss the limitations of the considered approach, caused by (i) the amplified spontaneous emission of the active medium in the case of a very weak seeding X-ray field, and (ii) reduction of the population difference at the lasing transition caused by the stimulated transitions in the case of a strong seeding field. In Sec. V we summarize the results. Finally, in the Appendix we derive the analytical solution, given in Sec. III.

## II. THEORETICAL MODEL

The basic set of equations, describing formation of attosecond X-ray pulses in the active medium of the hydrogen-like ions, dressed by the far-off-resonant IR field, was presented in [19]. For completeness and self-sufficiency of this paper we briefly describe it below.

Let us consider an active medium of hydrogen-like $C^{5+}$ ions with an inversion at the transition $n=1 \leftrightarrow n=2$ (where $n$ is the principal quantum number) [12, 14-16, 19, 20]. The active medium is simultaneously irradiated by (i) a seeding linearly polarized X-ray field, and (ii) a modulating optical field, which propagate along the same direction ($x$-axis) and have the same linear po-larization (along *z*-axis). At the entrance to the medium, $x=0$, the X-ray field is quasi-monochromatic and has a form

$$\vec{E}(x=0,t) = \frac{1}{2}\vec{z}_0 \tilde{E}_{inc}(t)\exp(-i\omega_{inc}t) + \text{c.c.}, \tag{1}$$

where $\vec{z}_0$ is a unit polarization vector, c.c. stands for complex conjugation, $\omega_{inc}$ is the carrier frequency of the field, which is close to the frequency of the transition $n=1 \leftrightarrow n=2$, and $\tilde{E}_{inc}(t)$ is the slowly varying amplitude of the X-ray field. For the analytical solution, derived below, we assume $\tilde{E}_{inc}(t)$ in the form of a unit step function, which is turned on instantly at the initial moment of time, $t=0$. In the numerical calculations, we imply an incident field with a rectangular envelope $\tilde{E}_{inc}(t)$ and smoothed turn-on and turn-off. The modulating optical field inside the medium has a form

$$\vec{E}_\Omega(x,t) = \vec{z}_0 E_M \cos\left[\Omega\left(t - \frac{n_{pl}}{c}x\right)\right]. \tag{2}$$

Here $E_M$ is the amplitude of the modulating field, $\Omega$ is its angular frequency, $c$ is the speed of light in vacuum, $n_{pl} = \sqrt{1-\omega_{pl}^2/\Omega^2}$ is the plasma refraction index at the frequency of the modulating field, $\omega_{pl} = \sqrt{4\pi N_e e^2/m_e}$ is the (electron) plasma oscillation frequency, $N_e$ is the concentration of free electrons; $e$ and $m_e$ are charge and mass of an electron, respectively. Eq. (2) implies that the modulating field is monochromatic and passes through the medium at the phase velocity $c/n_{pl}$. These approximations are justified if (i) the duration of the modulating optical field considerably exceeds both the duration of the X-ray field, and the relaxation times of the active medium; (ii) the concentration of free electrons is nearly constant in the interaction volume during the interaction time, and (iii) the modulating field is far-detuned from all the dipole-allowed transitions involving the populated states of the ions (so it is nor absorbed, neither is affected by the resonant dispersion).

The relevant energy level scheme, which describes the interaction of the resonant X-ray field (1) with hydrogen-like ions, includes the five states of the ions, namely, the ground state $|1\rangle=|1s\rangle$, corresponding to the energy level $n=1$, and the excited states $|2\rangle=(|2s\rangle+|2p,m=0\rangle)/\sqrt{2}$, $|3\rangle=(|2s\rangle-|2p,m=0\rangle)/\sqrt{2}$, $|4\rangle=|2p,m=1\rangle$, and $|5\rangle=|2p,m=-1\rangle$, which correspond to the energy level $n=2$. Here $m$ is a projection of the orbital momentum of the ions on the polarization direction of the modulating optical field (*z*-axis). Under the action of the modulating field (2) the upper resonant energy level of the ions is split into three sublevels because of the Stark effect. The two of them, which correspond to the states $|2\rangle$ and $|3\rangle$, oscillate in space and time along with the electric field strength of the modulating field due to the linear Stark effect, and also experience a relatively small time-independent shift due to the quadratic Stark effect. The third sublevel corresponds to the states $|4\rangle$ and $|5\rangle$, which remain degenerate and experience only a quadratic Stark shift. The resonant polarization of the medium has a form

$$\vec{P}(x,t) = N_{ion}\left[\vec{d}_{12}\rho_{21} + \vec{d}_{13}\rho_{31} + \vec{d}_{14}\rho_{41} + \vec{d}_{15}\rho_{51} + \text{c.c.}\right], \tag{3}$$

where $N_{ion}$ is density of the resonant ions, $\vec{d}_{1i}$ are electric dipole moments of the transitions $|i\rangle \leftrightarrow |1\rangle$, $i=2,3,4,5$, while $\rho_{i1}$ are the quantum coherencies at these transitions. The values of dipole moments $\vec{d}_{1i}$ in Eq. (3) are $\vec{d}_{12} = \vec{z}_0 d_{tr}$, $\vec{d}_{13} = -\vec{z}_0 d_{tr}$, $\vec{d}_{14} = \vec{d}_{15} = i\vec{y}_0 d_{tr}$, where $d_{tr} = \frac{2^7}{3^5 Z}ea_0$, $Z$ is nucleus charge of the ions ($Z$=6 for $C^{5+}$), and $a_0$ is the Bohr radius. The non-zero elements of

the matrix of dipole moments are also $\vec{d}_{22} = \vec{z}_0 d_{av}$ and $\vec{d}_{33} = -\vec{z}_0 d_{av}$, where $d_{av} = \frac{3}{Z} e a_0$. As it was discussed in [15], the incident z-polarized X-ray field (1) is amplified and generates the sidebands due to interaction with the transitions $|2\rangle \leftrightarrow |1\rangle$ and $|3\rangle \leftrightarrow |1\rangle$, whose dipole moments are oriented along z-axis, whereas the transitions $|4\rangle \leftrightarrow |1\rangle$ and $|5\rangle \leftrightarrow |1\rangle$, whose dipole moments are perpendicular to z-axis, are a source of the amplified spontaneous emission (ASE), which is polarized along y-axis, and reduces the gain for the z-polarized X-ray field by increasing population of the ground state. In the following we consider a plasma, which consists of only the resonant hydrogen-like ions and free electrons (there is no ions of the other ionization degree), so that the ion concentration can be expressed through the electron concentration as $N_{ion} = N_e/(Z-1)$. The time-evolution of the quantum state of the ions is described by the density-matrix equations:

$$\begin{cases} \dfrac{\partial \rho_{11}}{\partial t} = \gamma_{11} \sum_{k=2}^{5} \rho_{kk} - i\left[\hat{H}, \hat{\rho}\right]_{11}, \\ \dfrac{\partial \rho_{ij}}{\partial t} = -\gamma_{ij} \rho_{ij} - i\left[\hat{H}, \hat{\rho}\right]_{ij}, \; i, j = \{1, 2, 3, 4, 5\}, \; ij \neq 11, \end{cases} \quad (4)$$

which take into account that the ground state $|1\rangle$ has an infinite lifetime. In Eqns. (4) $\gamma_{ij}$ are decay rates of the density matrix elements $\rho_{ij}$, while $\hat{H}$ is the Hamiltonian of the system. In the presence of both the resonant X-ray field and the modulating optical field it has a form

$$\hat{H} = \begin{pmatrix} \hbar\omega_1 & -E_z d_{tr} & E_z d_{tr} & -iE_y d_{tr} & -iE_y d_{tr} \\ -E_z d_{tr} & \hbar\omega_2(t,x) & 0 & 0 & 0 \\ E_z d_{tr} & 0 & \hbar\omega_3(t,x) & 0 & 0 \\ iE_y d_{tr} & 0 & 0 & \hbar\omega_4 & 0 \\ iE_y d_{tr} & 0 & 0 & 0 & \hbar\omega_5 \end{pmatrix}, \quad (5)$$

where

$$\begin{cases} \hbar\omega_1 = -\dfrac{m_e e^4 Z^2}{2\hbar^2}\left\{1 + \dfrac{9}{256} F_0^2\right\}, \\ \hbar\omega_2(t,x) = -\dfrac{m_e e^4 Z^2}{8\hbar^2}\left\{1 + \dfrac{21}{4} F_0^2 + 3F_0 \cos\left(\Omega\left[t - n_{pl} \dfrac{x}{c}\right]\right)\right\}, \\ \hbar\omega_3(t,x) = -\dfrac{m_e e^4 Z^2}{8\hbar^2}\left\{1 + \dfrac{21}{4} F_0^2 - 3F_0 \cos\left(\Omega\left[t - n_{pl} \dfrac{x}{c}\right]\right)\right\}, \\ \hbar\omega_4 = \hbar\omega_5 = -\dfrac{m_e e^4 Z^2}{8\hbar^2}\left\{1 + \dfrac{39}{8} F_0^2\right\} \end{cases} \quad (6)$$

are the energies of the states $|1\rangle$, $|2\rangle$, $|3\rangle$, $|4\rangle$, and $|5\rangle$, correspondingly, with taking into account the linear and quadratic Stark shifts induced by the modulating field (2) [21]; $F_0 = \left(\dfrac{2}{Z}\right)^3 \dfrac{E_M}{E_A}$ is a dimensionless amplitude of the modulating field, and $E_A = m_e^2 e^5/\hbar^4 \cong 5.14 \cdot 10^9$ V/cm is the atomic unit of electric field. In turn, the decay rates $\gamma_{ij}$ are determined by the following equations:

$$\gamma_{21} = \gamma_{31} = \gamma_{coll} + w_{ion}^{(2,3)}/2 + \Gamma_{rad}/2 \equiv \gamma_z, \quad (7)$$

$$\gamma_{41} = \gamma_{51} = \gamma_{coll} + w_{ion}^{(4,5)}/2 + \Gamma_{rad}/2 \equiv \gamma_y,$$

$$\gamma_{32} = \gamma_{coll} + w_{ion}^{(2,3)} + \Gamma_{rad}, \quad \gamma_{54} = \gamma_{coll} + w_{ion}^{(4,5)} + \Gamma_{rad},$$

$$\gamma_{42} = \gamma_{52} = \gamma_{43} = \gamma_{53} = \gamma_{coll} + w_{ion}^{(2,3)}/2 + w_{ion}^{(4,5)}/2 + \Gamma_{rad},$$

$$\gamma_{11} = \Gamma_{rad}, \quad \gamma_{22} = \gamma_{33} = w_{ion}^{(2,3)} + \Gamma_{rad}, \quad \gamma_{44} = \gamma_{55} = w_{ion}^{(4,5)} + \Gamma_{rad}.$$

In (7) $\gamma_{coll}$ is the collision broadening of the spectral lines, $\Gamma_{rad}$ is the radiative decay rate from each of the upper states of the ions, $|2\rangle$, $|3\rangle$, $|4\rangle$, or $|5\rangle$, to the ground state $|1\rangle$; while $w_{ion}^{(2,3)}$ and $w_{ion}^{(4,5)}$ are the ionization rates from the states $|2\rangle$ or $|3\rangle$, and $|4\rangle$ or $|5\rangle$, respectively, averaged over the period of the modulating field:

$$w_{ion}^{(2,3)} = \frac{m_e e^4 Z^2}{16\hbar^3}\sqrt{\frac{3F_0}{\pi}}\left[\frac{4}{F_0}e^3 + \left(\frac{4}{F_0}\right)^3 e^{-3}\right]e^{-\frac{2}{3F_0}},$$

$$w_{ion}^{(4,5)} = \frac{m_e e^4 Z^2}{4\hbar^3}\sqrt{\frac{3F_0}{\pi}}\left(\frac{4}{F_0}\right)^2 e^{-\frac{2}{3F_0}}$$

(8)

Since the modulating field should not ionize the active medium during the interaction time, it should be not too strong, so that both (i) the quadratic Stark shifts of the resonant energy levels and (ii) the ionization rates from them remain much smaller than the frequency of the modulating field, $\Omega$. It allows taking them into account as time-independent values [22, 23, 15], as it is implied by Eqs. (6) and (8). Under the same conditions cubic Stark effect and higher-order corrections to the Stark shift can be safely neglected.

In the following we assume the active medium in the form of a long cylinder of the length $L$ and radius $R \ll L$, with Fresnel parameter $F = \pi R^2/(\lambda_{21} L) \sim 1$ (here $\lambda_{21} = 2\pi c/\bar{\omega}_{21}$ is the wavelength of the resonant X-ray field, and $\bar{\omega}_{21}$ is the time-averaged frequency of the transition $|2\rangle \leftrightarrow |1\rangle$). In such a case, propagation of the X-ray field through the medium can be described by one-dimensional wave equation:

$$\frac{\partial^2 \vec{E}}{\partial x^2} - \frac{\varepsilon_{X-ray}}{c^2}\frac{\partial^2 \vec{E}}{\partial t^2} = \frac{4\pi}{c^2}\frac{\partial^2 \vec{P}}{\partial t^2}, \quad (9)$$

where $\vec{E} = \vec{z}_0 E_z + \vec{y}_0 E_y$ is a vector of the resonant field inside the medium (although the seeding field (1) is $z$-polarized, $y$-polarization appears in the medium because of the spontaneous emission at the transitions $|4\rangle \leftrightarrow |1\rangle$ and $|5\rangle \leftrightarrow |1\rangle$), $\vec{P}$ is a vector of the resonant polarization (3), and $\varepsilon_{X-ray} = 1 - 4\pi N_e e^2/(m_e \omega_{inc}^2) \simeq 1$ is a dielectric permittivity of the plasma for the X-ray field. The system of equations (3)-(9) describe the transformation of the resonant X-ray field (1) in the active medium of a hydrogen-like plasma-based X-ray laser irradiated by the modulating field (2). Let us change the independent variables $x, t \to x, \tau = t - x\sqrt{\varepsilon_{X-ray}}/c$ and seek for a solution of this system within the slowly varying amplitude approximation for both the X-ray field and the resonant polarization, and the rotating wave approximation for the density matrix elements:

$$\begin{cases}\vec{E}(x,\tau) = \frac{1}{2}\{\vec{z}_0 \tilde{E}_z(x,\tau) + \vec{y}_0 \tilde{E}_y(x,\tau)\}e^{-i\omega_{inc}\tau} + c.c., \\ \vec{P}(x,\tau) = \frac{1}{2}\{\vec{z}_0 \tilde{P}_z(x,\tau) + \vec{y}_0 \tilde{P}_y(x,\tau)\}e^{-i\omega_{inc}\tau} + c.c.,\end{cases}$$

(10a)

$$\begin{cases} \rho_{i1}(x,\tau) = \tilde{\rho}_{i1}(x,\tau) e^{-i\omega_{inc}\tau}, \ i = \{2,3,4,5\}, \\ \rho_{ij}(x,\tau) = \tilde{\rho}_{ij}(x,\tau), \ ij \neq \{21,31,41,51\}, \\ \tilde{\rho}_{ij}(x,\tau) = \tilde{\rho}_{ji}^{*}(x,\tau) \end{cases} \quad (10b)$$

where $\tilde{E}_p$, $\tilde{P}_p$ ($p = z, y$) and $\tilde{\rho}_{ij}$ ($i, j = 1 \div 5$) are slow functions of space and time, which means

$$\frac{1}{|\tilde{E}_p|}\left|\frac{\partial \tilde{E}_p}{\partial x}\right|, \frac{1}{|\tilde{P}_p|}\left|\frac{\partial \tilde{P}_p}{\partial x}\right|, \frac{1}{|\tilde{\rho}_{ij}|}\left|\frac{\partial \tilde{\rho}_{ij}}{\partial x}\right| \ll \omega_{inc}\sqrt{\varepsilon_{X-ray}}/c \quad \text{and} \quad \frac{1}{|\tilde{E}_p|}\left|\frac{\partial \tilde{E}_p}{\partial \tau}\right|, \frac{1}{|\tilde{P}_p|}\left|\frac{\partial \tilde{P}_p}{\partial \tau}\right|, \frac{1}{|\tilde{\rho}_{ij}|}\left|\frac{\partial \tilde{\rho}_{ij}}{\partial \tau}\right| \ll \omega_{inc}.$$ Then, the equations (9), (4) take the form:

$$\begin{cases} \dfrac{\partial \tilde{E}_z}{\partial x} = i\dfrac{4\pi\omega_{inc}N_{ion}d_{tr}}{c\sqrt{\varepsilon_{X-ray}}}(\tilde{\rho}_{21} - \tilde{\rho}_{31}), \\ \dfrac{\partial \tilde{E}_y}{\partial x} = -\dfrac{4\pi\omega_{inc}N_{ion}d_{tr}}{c\sqrt{\varepsilon_{X-ray}}}(\tilde{\rho}_{41} + \tilde{\rho}_{51}). \end{cases} \quad (11)$$

The explicit form of the density matrix equations (4)-(6) under the considered approximations (10) is given in [15] and will not be repeated here because of its bulkiness. To solve the resulting system of equations (4)-(6), (11) one should have (i) the boundary conditions for the resonant field at the entrance of the medium, and (ii) the initial conditions for the density matrix elements.

Since $\varepsilon_{X-ray} \simeq 1$, the reflection of the incident X-ray field from the edges of the medium is negligible. Thus, the boundary conditions for the resonant field are

$$\tilde{E}_z(x=0,\tau) = \tilde{E}_{inc}(\tau), \quad \tilde{E}_y(x=0,\tau) = 0. \quad (12)$$

As initial conditions we assume that at $\tau = 0$ the ions are excited to the states $|2\rangle$-$|5\rangle$ with equal probability by a running wave of a pump laser field, which goes slightly ahead of the modulating optical field and the seeding X-ray field (while the rest of the ions remain in the ground state $|1\rangle$). Thus,

$$\tilde{\rho}_{11}(x,\tau=0) = \frac{1-4n_{tr}^{(0)}}{5}, \text{ and } \tilde{\rho}_{ii}(x,\tau=0) = \frac{1+n_{tr}^{(0)}}{5} \text{ for } i = \{2,3,4,5\}, \quad (13)$$

where $n_{tr}^{(0)}$ is the initial population difference at the transitions $|i\rangle \leftrightarrow |1\rangle$, $i = \{2,3,4,5\}$. At the same time, we account for the nonzero initial values of the quantum coherencies at the inverted transitions, $\tilde{\rho}_{i1}(x,\tau=0)$, which are responsible for spontaneous emission of the active medium [24]. In accordance with [19] these quantum coherencies are stepwise-constant random functions of the longitudinal coordinate, $x$, while the other coherencies equal zero:

$$\tilde{\rho}_{i1}(x_{k-1} \leq x < x_k, \tau=0) = A_{i,k}\frac{\exp\{i(\varphi_{i,k} + \phi_i)\}}{2N_{ion}\pi R^2 l_{elem}}, \ i = \{2,3,4,5\}, \quad (14a)$$

$$\tilde{\rho}_{ij}(x,\tau=0) = 0, \ i \neq j, \ i, j \neq 1, \quad (14b)$$

where $x_k = kl_{elem}$, $k = 1, 2,..., k_{max}$; $l_{elem}$ is much larger than the wavelength of the resonant X-ray field, $\lambda_{21} = 2\pi c/\bar{\omega}_{21}$, and much smaller than the total length of the medium, $\lambda_{21} \ll l_{elem} \ll L$. In Eqns. (14a) $\phi_2 = 0$, $\phi_3 = \pi$, $\phi_4 = \phi_5 = \pi/2$, while the amplitudes $A_{i,k}$ and phases $\varphi_{i,k}$ are random and statistically independent values, which obey the following probability distributions:

$$W(A_{i,k}^2) = \frac{1}{N_{ik}} \exp(-A_{i,k}^2/N_{i,k}), \quad 0 \le A_{i,k}^2 < \infty, \tag{15a}$$

$$W(\varphi_{i,k}) = 1/2\pi, \quad 0 \le \varphi_{i,k} < 2\pi. \tag{15b}$$

Here $N_{ik} = \rho_{ii}^{(0)} N_{ion} \pi R^2 l_{elem}$ is the number of particles, which are initially excited to the state $|i\rangle$ in the slice number $k$.

In order to get an insight into the process of sub-fs pulse formation from the quasi-monochromatic seeding X-ray field, in the next Section we derive a simplified analytical solution, which is further compared to the results of numerical calculations based on equations (4)-(6), (11) with boundary conditions (12) and initial conditions for the density matrix elements (see Eq. (13) and Appendix).

## III. ANALYTICAL STUDY

The high frequency of transitions $|i\rangle \leftrightarrow |1\rangle$, $i = \{2,3,4,5\}$, of $C^{5+}$ plasma-based X-ray laser results in a small radiative lifetime of the excited states $|i\rangle$, $\Gamma_{rad}^{-1} \simeq 1.23\,\text{ps}$, and rapid decrease of the population inversion at these transitions, $n_{i1} \equiv \tilde{\rho}_{ii} - \tilde{\rho}_{11}$, due to spontaneous emission. In the absence of the X-ray field inside the medium with initial conditions (13) the population differences at all the inverted transitions are identical, $n_{i1} = n_{tr}^{Sp}(\tau)$ for any $i = \{2,3,4,5\}$, where

$$n_{tr}^{Sp}(\tau) = (1 + n_{tr}^{(0)}) \exp\{-\Gamma_{rad}\tau\} - 1 \tag{16}$$

is the common population difference, whose time dependence is governed by radiative decay (spontaneous emission from) of the excited states. As follows from Eq. (16), the population difference is zero at the moment of time $\tau_0 = \frac{1}{\Gamma_{rad}} \ln(1 + n_{tr}^{(0)})$. In the following, similarly to [15], we assume that at the initial moment of time there is a complete population inversion, i.e. $n_{tr}^{(0)} = 1/4$. Besides, we assume that the other parameters of the active medium are also the same as in [15]. Thus, the densities of $C^{5+}$ ions and free electrons are $N_{ion} = 10^{17}\,\text{cm}^{-3}$ and $N_e = 5 \times 10^{17}\,\text{cm}^{-3}$, respectively, the ion temperature is 3 eV, the electron temperature is 5 eV, the intensity of the modulating optical field is $I_{opt} = 2.3 \times 10^{16}\,\text{W/cm}^2$ (which is slightly below the threshold of rapid ionization from the excited states of the ions). In such a case the population inversion at all the transitions $|2\rangle,|3\rangle,|4\rangle,|5\rangle \leftrightarrow |1\rangle$ drops to zero at $\tau = \tau_0 \simeq 270\,\text{fs}$, while the relaxation time of the quantum coherencies at the transitions $|2\rangle,|3\rangle \leftrightarrow |1\rangle$, which govern the amplification of z-polarized X-ray field, is $1/\gamma_z \simeq 415\,\text{fs}$ (the collision time is $1/\gamma_{coll} \simeq 560\,\text{fs}$ [25] and the inverse ionization rates from the upper lasing states are $1/w_{ion}^{(2,3)} \simeq 2.3\,\text{ps}$). Thus, $\tau_0 < 1/\gamma_z$, so that there is no steady-state solution for the quantum coherencies and the X-ray field inside the medium, contrary to the case of $Li^{2+}$ X-ray laser [19], where $\tau_0 \gg 1/\gamma_z$. In such a case, time-variation of the population differences should be taken into account in the analytical solution. To construct such an analytical solution, we assume that (i) the seeding z-polarized X-ray field is strong enough, so that the influence of y-polarized ASE of the active medium is negligible, and, at the same time, (ii) the spontaneous radiative transitions from the excited states of the active medium dominate over the stimulated transitions, so that the influence of the X-ray field on the population differences at the inverted transitions can be neglected. There is no contradiction between these two conditions for a sufficiently elongated active medium, since the population of the excited states is reduced because of spontaneous emission in any direction (in $4\pi$ solid an-gle), while the ASE originates only from the spontaneous emission along the axis of the active medium (in a small solid angle $\sim \pi R^2/L$). Within these approximations, the propagation of $z$-polarized X-ray field inside the medium is governed by the equations

$$\begin{cases} \dfrac{\partial \tilde{E}_z}{\partial x} = i \dfrac{4\pi \omega_z N d_{tr}}{c\sqrt{\varepsilon_{X-ray}}} (\tilde{\rho}_{21} - \tilde{\rho}_{31}), \\ \dfrac{\partial \tilde{\rho}_{21}}{\partial \tau} + \left( \gamma_z - i\Delta_\Omega \cos\left(\Omega \tau + \left(\sqrt{\varepsilon_{X-ray}} - n_{pl}\right)\dfrac{\Omega}{c} x\right)\right)\tilde{\rho}_{21} = -\dfrac{i}{2\hbar} n_{tr}^{Sp}(\tau) d_{tr} E_0(x,\tau), \\ \dfrac{\partial \tilde{\rho}_{31}}{\partial \tau} + \left( \gamma_z + i\Delta_\Omega \cos\left(\Omega \tau + \left(\sqrt{\varepsilon_{X-ray}} - n_{pl}\right)\dfrac{\Omega}{c} x\right)\right)\tilde{\rho}_{31} = \dfrac{i}{2\hbar} n_{tr}^{Sp}(\tau) d_{tr} E_0(x,\tau), \end{cases} \quad (17)$$

where the population differences at the transitions $|2\rangle \leftrightarrow |1\rangle$ and $|3\rangle \leftrightarrow |1\rangle$ are the functions of local time (16), and $\Delta_\Omega = \dfrac{3m_e e^4 Z^2}{8\hbar^3} F_0$ is the depth of modulation of frequencies of these transitions by the optical field due to the linear Stark effect.

To derive a simple analytical solution we assume, that the incident field (1) is turned on instantly at time $\tau|_{x=0} = t = 0$ and has constant amplitude after that:

$$\tilde{E}_z(x=0,\tau) = \theta(\tau) E_{inc}, \quad (18)$$

where $\theta(\tau)$ is Heaviside step function: $\theta(\tau) = 0$ for $\tau < 0$, and $\theta(\tau) = 1$ for $\tau \geq 0$.

In the following we assume that the incident X-ray field (1) is strictly resonant to the transitions $|2\rangle \leftrightarrow |1\rangle$ and $|3\rangle \leftrightarrow |1\rangle$ whose redefined frequencies include the time-averaged quadratic Stark shift, that is

$$\omega_{inc} = \omega_{21} = \omega_{31} = \dfrac{3m_e e^4 Z^2}{8\hbar^3}\left(1 - \dfrac{109}{64} F_0^2\right) \equiv \omega_z. \quad (19)$$

Let us look for a solution of Eqns. (17) for the envelope of the X-ray field in the form of a spectral comb

$$\tilde{E}_z(x,\tau) = \sum_{l=-\infty}^{\infty} E_l(x,\tau) e^{-i2l\Omega\tau}, \quad (20)$$

where $E_l(x,\tau)$ are slowly-varying functions of space and time on the time scale $\sim 2\pi/\Omega$ and the propagation distance $\sim 2\pi c/\Omega$ (then, the solutions for the quantum coherences have a similar form). Finally, let us suppose, that the quantum coherences $\tilde{\rho}_{21}$ and $\tilde{\rho}_{31}$ are determined only by the central spectral component of the X-ray field (at the frequency of the seeding radiation), which dominates over the sidebands at arbitrary time and arbitrary propagation distance inside the medium, $|E_0(x,\tau)| \gg |E_l(x,\tau)|$. As shown below, this is a rather good approximation in the case of $C^{5+}$ X-ray laser because of (i) not so high gain for the X-ray field and (ii) strong plasma dispersion for the modulating optical field. As shown in Appendix, under the above approximations the amplitudes of the spectral components of $z$-polarized X-ray field (20), which satisfy equations (17) and boundary condition (12), are:

$$E_0(x,\tau) = E_{inc}\theta(\tau)\exp\{g(P_\Omega,\tau)x\}, \quad (21a)$$

$$E_l(x,\tau) = E_{inc}\theta(\tau)\dfrac{J_{2l}(P_\Omega)}{J_0(P_\Omega)} g(P_\Omega,\tau) \dfrac{\exp\{(g(P_\Omega,\tau) - i2l\Delta K)x\} - 1}{g(P_\Omega,\tau) - i2l\Delta K}, \; l \neq 0, \quad (21b)$$

$$g(P_\Omega, \tau) = g_0 J_0^2(P_\Omega) \left[ \frac{1 + n_{tr}^{(0)}}{n_{tr}^{(0)}} \cdot \frac{e^{-\Gamma_{rad}\tau} - e^{-\gamma_z\tau}}{1 - \Gamma_{rad}/\gamma_z} - \frac{1}{n_{tr}^{(0)}}\left(1 - e^{-\gamma_z\tau}\right) \right], \qquad (21c)$$

where $g_0 = \dfrac{4\pi\omega_z n_{tr} d_{tr}^2 N_{ion}}{\hbar c \sqrt{\varepsilon_{X-ray}} \gamma_z}$ is the gain coefficient for the resonant X-ray field in the absence of the modulating field, and $\Delta K = \left(\sqrt{\varepsilon_{X-ray}} - n_{pl}\right)\Omega/c$ is a modification of the wavenumber of the modulating field due to the plasma dispersion. Thus, the amplified X-ray field consists of the resonant spectral component (21a), and a set of sidebands (21b) at the combinational frequencies, separated from the resonance by even multiples of the frequency of the modulating field. The resonant component of the X-ray field exponentially grows during its propagation through the medium with the effective gain coefficient (21c), which is a nonmonotonic function of local time; while the sidebands experience also an increasing phase-shift caused by mismatch of the phase velocities of the amplified X-ray field and the modulating optical field. A comparison of the derived analytical solution (21) with a similar solution for the case of $Li^{2+}$ X-ray laser (see Eqns. (19) in [19]) shows that an account of the local-time dependence of the population differences (14) changes only the local-time dependence of the effective gain coefficient $g(P_\Omega, \tau)$ for the X-ray field. Moreover, in the limit of slow radiative decay of the excited states (or fast relaxation of the quantum coherencies at the inverted transitions), $\Gamma_{rad} \ll \gamma_z$, the effective gain coefficient takes exactly the same form, as in [19], - $g(P_\Omega, \tau) = J_0^2(P_\Omega)\left(1 - e^{-\gamma_z\tau}\right)g_0$. Therefore, the spatial dependences of spectral components of the amplified X-ray field are the same as in the case of $Li^{2+}$ ions [19]. In particular, in the limit of a small gain / strong plasma dispersion, $g(P_\Omega, \tau) \ll |l|\Delta K$ (where $l$ is a sideband number), which is realized either at the initial moments of time, $\tau \ll \tau_0$, or for a low-frequency modulating field, $\Omega \sim \omega_{pl}$, or for the modulation indices, which satisfy the condition $J_0(P_\Omega) \approx 0$, the sidebands are much weaker as compared to the resonant spectral component of the X-ray field. In such a case, the amplitude of $l$-th sideband periodically oscillates in space, reaching the maximum values at odd multiples of a coherence length

$$L_{coh}^{(l)} = \frac{\lambda_{opt}}{4|l|\left(\sqrt{\varepsilon_{X-ray}} - n_{pl}\right)}, \qquad (22)$$

where $\lambda_{opt}$ is a wavelength of the modulating optical field. With increasing effective gain coefficient the sidebands become stronger, while their spatial oscillations become accompanied by the exponential grows, see Fig. 1, which is plotted for the value of modulation index $P_\Omega = 4.1$.

At the same time, contrary to the case of $Li^{2+}$ ions [19], a decrease of the population difference at the inverted transitions (16) with increasing local time results in nonmonotonic local-time dependence of both (i) the effective gain coefficient (21c) and (ii) the amplitudes of spectral components of the amplified X-ray field, (21a), (21b), shown in Fig. 2. The effective gain coefficient $g(P_\Omega, \tau)$ reaches its maximum value at the moment of time

$$\tau_{max} = \frac{1}{\gamma_z} \cdot \frac{1}{1 - \Gamma_{rad}/\gamma_z} \ln\left[\frac{\gamma_z}{\Gamma_{rad}}\left(1 - \frac{1 - \Gamma_{rad}/\gamma_z}{1 + n_{tr}^{(0)}}\right)\right], \qquad (23)$$

which is determined from the condition $dg(\tau)/d\tau\big|_{\tau=\tau_{max}} = 0$. For the considered parameters of the active medium $\tau_{max} \simeq 208$ fs. Such a location of a maximum in the time-dependence of the effective gain coefficient in Fig. 2(a) can be understood in the following way. The active medium has a finite response time to the seeding X-ray field, which is switched on instantly at $\tau = 0$. This response time is $1/\gamma_z \simeq 415\,\text{fs}$, so, the resonant response of the ions is intensified with increasing time till $\tau \leq 1/\gamma_z$. On the other hand, radiative decay of the excited states results in decrease of the population difference at the resonant transitions (16), thus weakening the resonant interaction between the X-ray field and the ions. At $\tau = \tau_{\max} < \tau_0, 1/\gamma_z$ these two factors balance each other, which results in maximum gain for the X-ray field. With further increasing time, the effective gain coefficient decreases and finally changes the sign at $\tau_{switch} \simeq 475\,\text{fs}$, after which the medium becomes absorbing for the X-ray field. Naturally, $\tau_{switch} > \tau_0$, since amplification or absorption of the X-ray field is governed by the values of quantum coherencies, rather than the population differences at the resonant transitions, and some time is needed for the coherencies (which possess considerably nonzero values at $\tau = \tau_0$), to change the sign after a change of sign of the population difference. In Fig. 2(b) we plot the local-time dependencies of the amplitudes of the resonant spectral component and the sidebands of the X-ray field predicted by the analytical solution (21) at the output from the active medium of the length $L = 3.9$ mm. The modulation index is $P_\Omega = 4.1$. As shown below, such a combination of $L$ and $P_\Omega$ corresponds to one of the optima for the transformation of a quasi-monochromatic seeding X-ray field (1) into an attosecond pulse train. As follows from Fig. 2(b), the amplitude of the resonant spectral component of the X-ray field reaches its maximum value along with the effective gain coefficient at $\tau = \tau_{\max}$, while at $\tau = \tau_{switch}$ (when the gain coefficient is zero), it equals the amplitude of the seeding field. At $\tau > \tau_{switch}$ the medium starts to absorb the radiation, and the amplitude of the resonant spectral component of the X-ray field monotonically tends to zero. The amplitudes of the sidebands also reach their local maxima at $\tau = \tau_{\max}$, while at $\tau = \tau_{switch}$ they become zero. Until $\tau \leq \tau_{switch}$ the resonant spectral component of the X-ray field dominates over the sidebands, which justify the analytical solution (21). However, for $\tau > \tau_{switch}$ the sidebands grow up again due to the energy transfer from the resonant spectral component in the absorbing medium, see [23] and [26]. In such a case, the amplitudes of the sidebands may become comparable or even exceed the amplitude of the resonant spectral component of the field [27], which makes the solution (21) inapplicable (in particular, the amplitudes of sidebands will never exceed the amplitude of the seeding field, as it happens in Fig. 2(b) for $\tau \geq 850\,\text{fs}$, because of the energy dissipation in absorbing medium). Thus, for the considered parameters of the plasma, the analytical solution (21) is valid only till the medium amplifies the resonant radiation, $\tau \leq \tau_{switch}$, or for a little bit longer time interval (for $\tau \leq 550\,\text{fs}$ in the considered case).

In Fig. 3 we plot the maximum achievable values of the amplitudes of sidebands, normalized to the amplitude of the resonant spectral component of the X-ray field, calculated via the analytical solution (21), as a function of the modulation index. Generally, the peak amplitudes of sidebands, shown in Fig. 3, are achieved at different depths of the active medium for each sideband number and each value of the modulation index (but at the same moment of time $\tau = \tau_{\max}$). For each point in Fig. 3 the amplitudes of sidebands are normalized to the amplitude of the resonant spectral component at the corresponding propagation distance (where the particular sideband reaches its maximum value). As follows from this figure, the amplitudes of sidebands never reach the amplitude of the resonant spectral component of the X-ray field (at the frequency of the seed). With increasing modulation index, an increasing number of sidebands possess comparable amplitudes. Thus, for $0 < P_\Omega < 2.4$ only the resonant spectral component and the ±1 sidebands are appreciably nonzero; for $2.4 < P_\Omega < 5.5$ also the ±2 sidebands should be taken into account; for $5.5 < P_\Omega < 8.4$ the ±3 sidebands become noticeable, and so on. But at the same time, with increasing value of the modulation index, the amplitudes of sidebands become smaller with respect to the amplitude of the resonant spectral component. In order to constitute a train of attosecond pulses, the spectral components of the output X-ray field should (i) be phase-matched and (ii) have comparable amplitudes. Thus, the resonant spectral component of the X-ray field should be attenuated to the level of sidebands. Use of larger modulation indices allows for generation of a broader spectrum (and thus supports formation of shorter pulses with higher off-duty ratio), but requires stronger attenuation of the resonant spectral component of the output X-ray field, which results in lower energy and lower peak intensity of the pulse train. The possibilities for the pulse formation using different modulation indices and different attenuation levels for the central spectral component of the output X-ray field are considered in the next Section.

## IV. X-RAY PULSE FORMATION IN HYDROGEN-LIKE PLASMA OF $C^{5+}$

In this section we will analyze the optimal conditions for the attosecond pulse formation from the resonant X-ray field on the basis of both the derived analytical solution (21), and the numerical solution of Eqns. (4)-(6), (11). During the propagation through the medium the spectrum of the resonant field is enriched by multiple sidebands. Under certain conditions, these sidebands are sufficiently strong and nearly phase-aligned, so that their constructive interference in time-domain results in formation of an attosecond pulse train.

The X-ray field at the output from the medium (its z-polarization component, which corresponds to the amplified seeding field (1)) can be characterized by a contrast, $C$, which is a difference between the maximum and minimum values of its intensity within a half-cycle of the modulating field, normalized to the mean value of this intensity, averaged over the same time-interval:

$$C(x, P_\Omega, \tau) = \left(\max\{I_z\} - \min\{I_z\}\right) / \mathrm{mean}\{I_z\}, \qquad (24)$$

where $I_z = \frac{c}{8\pi}\left|\tilde{E}_z(x, P_\Omega, \tau)\right|^2$. For the seeding field (18) with the time-independent amplitude the contrast equals unity, while the formation of the pulse train results in $C \gg 1$. The higher contrast generally corresponds to the higher off-duty ratio and better pulse shape. Thus, the optimal conditions for the pulse formation correspond to the maxima in the dependence of contrast on the length of the medium, $x$, and the modulation index, $P_\Omega$. The dependence $C = C(x, P_\Omega)$, calculated via the analytical solution (21) at the peak of the intensity envelope (in the vicinity of $\tau = \tau_{\max} \approx 208\,\mathrm{fs}$) is shown in Fig. 4, panel (a). The peak contrast, $C \approx 2.8$, is achieved at $x=3.9$ mm and $P_\Omega = 3.2$, which maximize the amplitudes of ±1 sidebands (which dominate over the rest of sidebands, but still remain weaker than the resonant spectral component), see Fig. 3. The higher values of contrast can be achieved via external attenuation of the central spectral component of the X-ray field either by a resonant absorber or via the spectrally-selective X-ray optics [28, 29]. Panels (b), (c), and (d) of Fig. 4 show the contrast of X-ray field after attenuation of the amplitude (not spectral density) its central spectral component by the factors $K = 2$, $K = 4$, and $K = 8$, respectively. As can be seen, with increasing attenuation factor the peak values of contrast grow: $C \approx 3.5$ in (b), $C \approx 4.6$ in (c), and $C \approx 5.2$ in (d); the optimal value of the modulation index increases: $P_\Omega = 4.3$ in (b), $P_\Omega = 4.5$ in (c), and $P_\Omega = 4.7$ in (d); while the optimal propagation distance remains nearly constant: $x=3.3$ mm in (b) and (c), $x=3.4$ mm in (d). For the values of the modulation index $4.3 \leq P_\Omega \leq 4.7$ the amplitudes of ±1 and ±2 sidebands become comparable to each other (and the ±3 sidebands become noticeable), but the sidebands are weaker with respect to the resonant spectral component of the field as compared to the case of $P_\Omega = 3.2$ (which is the optimum in Fig. 4(a)), see Fig. 3. Thus, with increasing attenuation factor the spectrum of X-ray field is enriched by an increasing number of sidebands, but contains less energy.

The shape of individual pulses at the peak of the intensity envelope (at the vicinity of $\tau = \tau_{max} \approx 208$ fs) under the conditions, which maximize the contrast in the absence of attenuation ($K = 1$), as well as for the attenuation factors $K = 2$, $K = 4$, and $K = 8$, is shown in Fig. 5. As follows from this figure, with increasing attenuation factor the pulse duration comprises a smaller fraction of the modulating field cycle. At this point it is worth noting that in Fig. 4 and Fig. 5 the intensity of the modulating field is fixed to the value $I_{opt} = 2.3 \times 10^{16}$ W/cm$^2$, which slightly below the threshold of rapid ionization from the excited states of the active medium (so that the further increase of intensity is not possible). Such a high intensity of the modulating field allows to produce the pulses with highest contrast and shortest duration. At the same time, the pulses can be produced using longer wavelength modulating field of lower intensity (as compared to those discussed below). In the considered case, the modulation index is changed via a change of wavelength of the modulating field: a larger value of the modulation index corresponds to a larger wavelength. In Fig. 5 the wavelengths of the modulating field are: 357.8 nm for $K = 1$, 480.8 nm for $K = 2$, 503.2 nm for $K = 4$, and 525.5 nm for $K = 8$. Thus, in each case the duration of the modulating field cycle is different. The relative full width at half maximum (FWHM) duration of the pulses in the modulating field cycles, $2\pi/\Omega$, is 0.182 for $K = 1$, 0.115 for $K = 2$, 0.092 for $K = 4$, and 0.076 for $K = 8$. In absolute values the pulse duration is 217 as for $K = 1$, 184 as for $K = 2$, 154 as for $K = 4$, and 133 as for $K = 8$.

At the same time, the peak intensity of the pulses decreases with increasing attenuation factor: $I_{max} = 23.1 I_0$ for $K = 1$, $I_{max} = 12.1 I_0$ for $K = 2$, $I_{max} = 3.67 I_0$ for $K = 4$, and $I_{max} = 1.05 I_0$ for $K = 8$. These values are predicted by the analytical solution (21), $I_0$ is the intensity of the seeding field (1). Thus, an increase of the attenuation factor by two times results in 2–3 times decreasing pulse peak intensity. The intensity of the pulses can be increased in three ways: (i) by choosing a smaller attenuation factor, $K$, (ii) via an increase of the propagation distance, $x$, and (iii), in the case of attenuation factors $K \geq 2$, via a reduction of the modulation index, $P_\Omega$. The role of the attenuation factor is trivial. An increase of the propagation distance (the medium length) results in increasing gain-length product and stronger amplification of the X-ray field. In its turn, a decrease of the modulation index from $4.3 \leq P_\Omega \leq 4.7$ to $3.2 \leq P_\Omega \leq 4.2$ results in increasing amplitudes of ±1 and ±2 sidebands at the cost of a larger difference between them and weaker ±3 sidebands. A decrease of the modulation index below $P_\Omega = 3.2$ is not favorable since it results in increasing amplitude of the central spectral component of the X-ray field and growing pulse pedestal.

In the following we will consider the case of $x=3.9$ mm, $P_\Omega = 4.1$ (the wavelength of the modulating field is 458.4 mn), and the attenuation factor $K = 2$ (see Fig. 1 and Fig. 2). According to the analytical solution, these parameters correspond to $C \approx 3.25$ and the peak intensity of the pulse train $I_{max} = 26.5 I_0$. Below we present the results of numerical solution of Eqns. (4)-(6), (11) for this case. Looking ahead, it is worth noting that the results of numerical calculations for the dependence of contrast of the X-ray field, $C$, on the propagation distance, $x$, and the modulation index, $P_\Omega$, strongly resembles those shown in Fig. 4 except for the values of modulation index $P_\Omega \approx 2.4, 5.5, 8.6$, where $J_0(P_\Omega) = 0$ and the analytical solution become inapplicable, while for these values of the modulation index the output intensity is quite small, so that they are of minor importance. Thus, the analytical solution correctly predicts the optimal conditions for the pulse formation. In its turn, the numerical solution describes the pulse formation more precisely by taking into account the rescattering of sidebands into each other, their nonlinear interaction with the medium (in particular, reduction of the population differences because of the stimulated transitions) and the amplified spontaneous emission of y-polarization. In order to directly compare the results of the calculations with the analytical solution derived in the previous section, we assume an incident X-ray field with a rectangular shape and smoothed turn-on and turn-off:

$$\tilde{E}_{inc}(t) = E_0 \times \begin{cases} \sin^2\left(\dfrac{\pi}{2}\dfrac{t}{t_{switch}}\right), & 0 \leq t < t_{switch}, \\ 1, & t_{switch} \leq t < t_{flat} + t_{switch}, \\ \cos^2\left(\dfrac{\pi}{2}\dfrac{\left[t - \{t_{flat} + t_{switch}\}\right]}{t_{switch}}\right), & t_{flat} + t_{switch} \leq t < t_{flat} + 2t_{switch}, \\ 0, & t \geq t_{flat} + 2t_{switch}, \end{cases} \quad (25)$$

where $t_{flat}$ = 1470 fs and $t_{switch}$ = 15 fs, which is much longer than the X-ray field cycle, $2\pi/\omega_{inc}$ ≈ 11 as. The wavelength of the seeding field is $2\pi c/\omega_{inc}$ = 3.376 nm. The radius of the plasma channel is 2.5 μm. The time-dependence of intensity of the output X-ray field for these parameter values and the intensity of the seeding field $I_0 = 10^9$ W/cm$^2$ is shown in Fig. 6 (the central spectral component of the X-ray field is attenuated two times). At the peak of the intensity envelope the pulse shape is nearly the same as in the optimal case, shown in Fig. 5 for $K = 2$; the pulse duration is 180 as, the contrast of the field is 3.2, and the peak intensity of the pulses is $I_{max} \approx 10 I_0$. The peak intensity of the pulses resulting from the numerical calculations for is lower than that predicted by the analytical solution. There are two reasons for this. Firstly, the inertialess approximation, used for the derivation of the analytical solution (see Appendix), leads to overestimated value of the amplified X-ray field. Secondly, stimulated transitions, induced by the seeding field (and neglected by the analytical solution), result in faster depletion of the population inversion of the medium, and smaller amplification of the resonant field. However, the pulses are produced not only in the amplifying medium, but also after the medium becomes absorbing for the resonant X-ray field at $\tau \approx 460$ fs $\approx \tau_{switch}$. This is not surprising, since in accordance with [23] the value $P_\Omega = 4.1$ is within the optimal range of modulation indices for the pulse formation in the absorbing medium: $v_0^{(1)} < P_\Omega < v_2^{(1)}$, where $v_0^{(1)}$ is the first root of equation $J_0(P_\Omega) = 0$ and $v_2^{(1)}$ is the first root of equation $J_2(P_\Omega) = 0$. The duration of the pulses, produced in the absorbing medium is 130 as, the contrast of the field is 5.4, the pulse peak intensity is $1.7 I_0$. Noteworthy, that in this case the contrast is higher than in the case of amplifying medium and attenuation factor $K = 8$ (see Fig. 5), while the peak intensity is also higher. Thus, instead of using large attenuation factors, it might be more reasonable to wait till the medium becomes absorbing for the X-ray field. In Fig. 7 we plot the Fourier transform (amplitude spectrum) of the field, shown in Fig. 6. Fig. 7(a) shows the time-windowed spectrum, calculated via a convolution of the X-ray field with the time-window of the form $F(\tau') = \sin^2\{\pi[\tau' - (\tau - \tau_d)]/(2\tau_d)\}$ if $\tau - \tau_d \leq \tau' \leq \tau + \tau_d$, and $F(\tau') = 0$ otherwise; where $\tau_d$ = 20 fs is the full duration of the time-window at the half of its maximum, and $\tau$ is the position of the maximum, which varies from 0 to 1500 fs. Figs. 7(b) and 7(c) show the cuts of the surface, plotted in Fig. 7(a) for $\tau$ = 208 fs and $\tau$ = 1210 fs, respectively. Thus, Figs. 7(b) and 7(c) show the time-windowed spectra of the pulses, shown at the inserts of Fig. 6. As follows from Fig. 7(b), at the peak of the intensity envelope the amplitudes of ±2 spectral components are larger, than those predicted by the analytical solution (see Figs. 1 and 2), which can be attributed to the rescattering of sidebands. At the same time, in agreement with the analytical solution, the amplitudes of sidebands become close to zero at $\tau \approx 460$ fs $\approx \tau_{switch}$, see Fig. 7(a). In the absorbing medium (Fig. 7(c)) the amplitude of the central spectral component is smaller and closer to the amplitudes of sidebands, which is the reason for the high contrast of the pulses, produced in this case.

In Fig. 8 and Fig. 9 we compare the time-dependencies of intensity of the X-ray field, calculated numerically for different intensities of the seed (1): $I_0 = 10^6$ W/cm$^2$, $I_0 = 10^9$ W/cm$^2$, and $I_0 = 10^{12}$ W/cm$^2$, with that predicted by the analytical solution (all the other parameters are the same as in Figs. 6 and 7). Fig. 8 shows the envelopes of the pulse trains within the time interval $0 \leq \tau \leq 1485$ fs, while in Fig. 9 we compare the shape of individual pulses within a single cycle of the modulating field $2\pi/\Omega \approx 1.52$ fs. Particularly, in Fig. 9(a) we plot the pulses produced in the amplifying medium and located at the peaks of the intensity envelopes (for each solution the peak is achieved at a different time, 95 fs $\leq \tau \leq$ 215 fs), while in Fig. 9(b) we plot the pulses produced in the absorbing medium (after a change of sign of the population inversion) and located at the tails of the pulse trains (in this case, depending on the solution, 740 fs $\leq \tau \leq$ 1180 fs). Thus, in Fig. 9 different solutions are shown in different time windows of the same duration $2\pi/\Omega \approx 1.52$ fs (whose left boundaries are artificially shifted to zero time). Also, in Fig. 8 we plot the envelope of the (y-polarized) amplified spontaneous emission of the active medium, which is noticeable for $I_0 = 10^6$ W/cm$^2$, while for the higher intensities of the seeding field it can be neglected. In Fig. 8 the intensity of the X-ray field is normalized to the intensity of the seeding field, $I_0$, while in Fig. 9 for each solution the intensity is normalized by its peak value within the chosen time window. In accordance with Fig. 8, the peak intensities of the pulses, produced in the amplifying medium, are: $I_{max} = 13.6\ I_0$ for $I_0 = 10^6$ W/cm$^2$; $I_{max} = 10.3\ I_0$ for $I_0 = 10^9$ W/cm$^2$; and $I_{max} = 2.85\ I_0$ for $I_0 = 10^{12}$ W/cm$^2$; the analytical solution predicts $I_{max} = 26.5\ I_0$. The peak intensities of the pulses, produced in the absorbing medium (after a change off sign of the population inversion), are: $I_{abs} = 1.98\ I_0$ for $I_0 = 10^6$ W/cm$^2$; $I_{abs} = 1.68\ I_0$ for $I_0 = 10^9$ W/cm$^2$; and $I_{abs} = 1.24\ I_0$ for $I_0 = 10^{12}$ W/cm$^2$. In Fig. 8 the peak intensity of the pulses, normalized to the intensity of the seeding field, $I_{max}/I_0$, decreases with increasing $I_0$ because of the stimulated emission of radiation, which reduces the population differences at the inverted transitions. The analytical solution overestimates the pulse peak intensity because of the use of inertialess approximation for the dependence of the resonant polarization of the medium on the electric field strength (see Appendix). At the same time, the relative intensity of the pulses, produced in the absorbing medium, $I_{abs}/I_0$, weakly depends on $I_0$, which makes possible (and reasonable) the use of absorbing medium for the pulse formation at intensities of the seeding field $I_0 > 10^{12}$ W/cm$^2$. The shape of the pulses in Fig. 9 also remains almost the same for different intensities of the seeding field, $I_0$, except for the highest considered intensity, $I_0 = 10^{12}$ W/cm$^2$, at which the generation of sidebands becomes less efficient because of the reduced population differences at the resonant transitions, which leads to increasing background of the pulses.

In summary, Figs. 8 and 9 show the possibility to produce an attosecond pulse train in a wide range of intensities of the seeding X-ray field, $I_0 = 10^6$-$10^{12}$ W/cm$^2$. At $I_0 = 10^6$ W/cm$^2$ the limiting factor is the amplified spontaneous emission, which takes a considerable part of the energy, initially stored in the population inversion of the active medium; while at $I_0 = 10^{12}$ W/cm$^2$ it is the stimulated emission of radiation, which reduces the population differences at the resonant transitions and limits the gain. For the considered parameters of the problem, the peak intensity of the pulses can reach $2.8 \times 10^{12}$ W/cm$^2$ for the pulse duration 180 as; even shorter 130 as pulses can be produced after a change of sign of the population inversion at the resonant transitions of the medium at the cost of a few times lower pulse peak intensity, $1.2 \times 10^{12}$ W/cm$^2$. The pulses are nearly transform-limited and the pulse shape weakly depends on the intensity of the seeding field.

## V. CONCLUSION

In this paper, we studied the ultimate capabilities and limitations for the attosecond pulse formation via optical modulation of an active medium of the hydrogen-like C$^{5+}$ plasma-based X-ray laser. The pulses are produced via the spectral broadening of a quasimonochromatic seeding X-ray field at a wavelength 3.38 nm in the "water window" range during its amplification by C$^{5+}$ ions, whose transition frequencies oscillate in time along with oscillation of the modulating optical field due to the linear Stark effect. This method was proposed in [15] on the bases of the numerical calculations. In the present paper we derived an analytical solution for the amplified (and spectrally broadened) X-ray field, which takes into account the plasma dispersion of the medium and rapid depopulation of the upper lasing states of $C^{5+}$ ions because of the radiative transitions. The analytical solution allowed us to find the optimal conditions for the transformation of a quasicontinuous seeding X-ray field into a train of attosecond pulses with the highest contrast and shortest duration. It has been shown that the contrast of pulses can be increased via external attenuation of the resonant spectral component of the amplified X-ray field to the level of the generated sidebands. The predictions of analytical theory have been compared to the results of numerical calculations, which take into account an amplified spontaneous emission of the active medium as well as a variety of nonlinear processes in the considered system. It has been shown, that the analytical solution correctly predicts the shape, duration and contrast of the attosecond pulses, as well as the envelope of the pulse train. At the same time, it overestimates the peak intensity of the pulses, which is reduced due to the final response time of the ions on the X-ray field and the stimulated emission at the lasing transitions. Besides, the numerical calculations show that the pulses continue to be formed after a change of sign of the population differences at the lasing transitions (when the medium becomes absorbing for the X-ray field). In this case the pulses are shorter and have better shape as compared to the amplifying medium at the cost of the lower peak intensity (which still can exceed the intensity of the seeding field). It has been shown, that the pulses can be produced from a seeding field with the intensity in the range $I_0 = 10^6\text{-}10^{12}$ W/cm$^2$. The peak intensity of the attosecond pulses can exceed the intensity of the seeding field by more than 10 times, while the pulses duration vary in the range 130-180 as. Such pulses at a wavelength 3.38 nm in the "water window" range might be a useful tool for the studies of the ultrafast processes in medicine and biology.

## AKNOWLEGMENTS

We appreciate helpful and stimulating discussions with Profs. Szymon Suckewer and Marlan Scully. We acknowledge support from Russian Foundation for Basic Research (RFBR, Grant No.18-02-00924), as well as support from National Science Foundation (NSF, Grant No. PHY-201-21-94), AFOSR (Grant No. FA9550-18-1-0141) and ONR (Grant No. N00014-16-1-3054). The analytical studies presented in Sections II-IV were supported by the Russia Science Foundation (RSF), Grant No. 16-12-10279. V.A.A. acknowledges support by the Foundation for the Advancement of Theoretical Physics and Mathematics BASIS.

## APPENDIX. DERIVATION OF THE ANALYTICAL SOLUTION

In order to solve Eqns. (17), let us use a substitution

$$\tilde{\rho}_{21}(x,\tau) = \hat{\rho}_{21}(x,\tau) e^{-\gamma_z \tau + iP_\Omega \sin\left[\Omega \tau + \left(\sqrt{\varepsilon_{X-ray}} - n_{pl}\right)\frac{\Omega}{c}x\right]} = $$
$$= \hat{\rho}_{21}(x,\tau) e^{-\gamma_z \tau} \sum_{k=-\infty}^{\infty} J_k(P_\Omega) e^{ik\Omega\tau} e^{ik\left(\sqrt{\varepsilon_{X-ray}} - n_{pl}\right)\frac{\Omega}{c}x}, \quad (A1)$$

where the Jacobi-Anger formula is used: $\exp\{iP_\Omega \sin(\varphi)\} = \sum_{k=-\infty}^{\infty} J_k(P_\Omega)\exp(ik\varphi)$, $J_k(P_\Omega)$ is Bessel function of the first kind of order $k$, and $P_\Omega = \Delta_\Omega / \Omega$ is modulation index, which is the amplitude of variation of frequencies of the transitions $|2\rangle \leftrightarrow |1\rangle$ and $|3\rangle \leftrightarrow |1\rangle$ due to linear Stark effect, normalized to the frequency of the modulating field. As follows from the second equation of the system (17), the function $\hat{\rho}_{21}(x,\tau)$ satisfies equation

$$\frac{\partial \hat{\rho}_{21}}{\partial \tau} = -i\frac{d_{tr}}{2\hbar} n_{tr}^{Sp}(\tau) E_0(x,\tau) \exp\left\{\gamma_z \tau - iP_\Omega \sin\left[\Omega\tau + \left(\sqrt{\varepsilon_{X-ray}} - n_{pl}\right)\frac{\Omega}{c}x\right]\right\}, \quad (A2)$$

which has a solution

$$\hat{\rho}_{21} = -i\frac{d_{tr}}{2\hbar} \sum_{m=-\infty}^{\infty} J_m(P_\Omega) \exp\left[-im\frac{\Omega}{c}\left(\sqrt{\varepsilon_{X-ray}} - n_{pl}\right)x\right] \int_0^\tau E_0(x,\tau') n_{tr}^{Sp}(\tau') e^{(\gamma_z - im\Omega)\tau'} d\tau' \quad (A3)$$

valid in the case $\tilde{\rho}_{21}(x,\tau=0)=0$. One can approximately evaluate the integral in the right-hand side of Eq. (A3) as

$$\int_0^\tau E_0(x,\tau') n_{tr}^{Sp}(\tau') e^{(\gamma_z - im\Omega)\tau'} d\tau' = \left\|\begin{matrix}\Delta\tau' = \tau - \tau' \\ d\Delta\tau' = -d\tau'\end{matrix}\right\| = e^{(\gamma_z - im\Omega)\tau} \int_0^\tau E_0(x, \tau - \Delta\tau') n_{tr}^{Sp}(\tau - \Delta\tau') e^{-(\gamma_z - im\Omega)\Delta\tau'} d\Delta\tau' \approx$$

$$\left\|n_{tr}^{Sp}(\tau) = \left(1+n_{tr}^{(0)}\right)e^{-\Gamma_{rad}\tau} - 1\right\| \approx e^{(\gamma_z - \Gamma_{rad} - im\Omega)\tau}\left(1+n_{tr}^{(0)}\right) E_0(x,\tau) \int_0^\tau e^{-(\gamma_z - \Gamma_{rad} - im\Omega)\Delta\tau'} d\Delta\tau' - \quad (A4)$$

$$-e^{(\gamma_z - im\Omega)\tau} E_0(x,\tau) \int_0^\tau e^{-(\gamma_z - im\Omega)\Delta\tau'} d\Delta\tau' = \left(1+n_{tr}^{(0)}\right) E_0(x,\tau) \frac{e^{(\gamma_z - \Gamma_{rad} - im\Omega)\tau} - 1}{\gamma_z - \Gamma_{rad} - im\Omega} - E_0(x,\tau) \frac{e^{(\gamma_z - im\Omega)\tau} - 1}{\gamma_z - im\Omega}$$

Formally, this approximation implies that the amplitude of the resonant spectral component of the field, $E_0(x,\tau)$, is a slowly-varying function of time at the time-intervals $\sim \gamma_z^{-1}$ and $\sim \Gamma_{rad}^{-1}$. But as follows from the derived solution, for the incident field with a time-independent amplitude (18), and for the values of time $\tau \leq \tau_{switch} \simeq 475\text{fs}$, Eq. (A4) gives a sufficiently good approximation. Thus, one finds a solution for $\tilde{\rho}_{21}(x,\tau)$ in the form

$$\tilde{\rho}_{21} = -i\frac{d_{tr}}{2\hbar} E_0(x,\tau) \sum_{m,k=-\infty}^{\infty} J_m(P_\Omega) J_k(P_\Omega) \exp\left[-i(m-k)\left(\sqrt{\varepsilon_{X-ray}} - n_{pl}\right)\frac{\Omega}{c}x\right] \times$$

$$\times \left\{\left(1+n_{tr}^{(0)}\right)\frac{e^{-\Gamma_{rad}\tau - i(m-k)\Omega\tau} - e^{ik\Omega\tau - \gamma_z\tau}}{\gamma_z - \Gamma_{rad} - im\Omega} - \frac{e^{-i(m-k)\Omega\tau} - e^{ik\Omega\tau - \gamma_z\tau}}{\gamma_z - im\Omega}\right\} \quad (A5)$$

The solution for $\tilde{\rho}_{31}(x,\tau)$ is found analogously, so that

$$\tilde{\rho}_{21} - \tilde{\rho}_{31} = -i\frac{d_{tr}}{2\hbar} E_0(x,\tau) \sum_{m,k=-\infty}^{\infty} \left\{1+(-1)^{m-k}\right\} J_m(P_\Omega) J_k(P_\Omega) \exp\left\{-i(m-k)\left(\sqrt{\varepsilon_{X-ray}} - n_{pl}\right)\frac{\Omega}{c}x\right\} \times$$

$$\times \left\{\left(1+n_{tr}^{(0)}\right)\frac{e^{-\Gamma_{rad}\tau - i(m-k)\Omega\tau} - e^{ik\Omega\tau - \gamma_z\tau}}{\gamma_z - \Gamma_{rad} - im\Omega} - \frac{e^{-i(m-k)\Omega\tau} - e^{ik\Omega\tau - \gamma_z\tau}}{\gamma_z - im\Omega}\right\} \quad (A6)$$

In the following we assume, that the frequency of the modulating field is much larger than the linewidth of the resonant transition, $\Omega/\gamma_z \gg 1$, so that for $J_0(P_\Omega) \neq 0$ a contribution with $m=0$ is dominant in Eq. (A6), which is reduced to

$$\tilde{\rho}_{21} - \tilde{\rho}_{31} = -i\frac{d_{tr} n_{tr}^{(0)}}{\hbar \gamma_z} E_0(x,\tau) \left\{\frac{1+n_{tr}^{(0)}}{n_{tr}^{(0)}} \frac{e^{-\Gamma_{rad}\tau} - e^{-\gamma_z\tau}}{1 - \Gamma_{rad}/\gamma_z} - \frac{1}{n_{tr}^{(0)}}\left(1 - e^{-\gamma_z\tau}\right)\right\} \times$$

$$\times J_0(P_\Omega) \sum_{l=-\infty}^{\infty} J_{2l}(P_\Omega) e^{-i2l\Omega\tau} \exp\left\{-i2l\left(\sqrt{\varepsilon_{X-ray}} - n_{pl}\right)\frac{\Omega}{c}x\right\} \quad (A7)$$

Thus, according to (18), z-polarized XUV field is determined by the equation

$$\frac{\partial \tilde{E}_z}{\partial x} = \frac{4\pi\omega_z N_{\text{ion}} n_{tr}^{(0)} d_{tr}^2}{\hbar c \sqrt{\varepsilon_{X-ray}} \gamma_z} \left\{ \frac{1+n_{tr}^{(0)}}{n_{tr}^{(0)}} \frac{e^{-\Gamma_{rad}\tau} - e^{-\gamma_z \tau}}{1 - \Gamma_{rad}/\gamma_z} - \frac{1}{n_{tr}^{(0)}} \left(1 - e^{-\gamma_z \tau}\right) \right\} \times$$

$$\times E_0(x,\tau) J_0(P_\Omega) \sum_{l=-\infty}^{\infty} J_{2l}(P_\Omega) e^{-i2l\Omega\tau} e^{-i2l\left(\sqrt{\varepsilon_{X-ray}} - n_{pl}\right)\frac{\Omega}{c}x}, \quad \text{(A8)}$$

which has a solution in the form of the spectral comb (20), that is $\tilde{E}_z(x,\tau) = \sum_{l=-\infty}^{\infty} E_l(x,\tau) e^{-i2l\Omega\tau}$.

The amplitudes of the spectral components of z-polarized field (20), which satisfy the boundary condition (18), are given by equations (21) of the paper.

*Corresponding author: antonov@appl.sci-nnov.ru

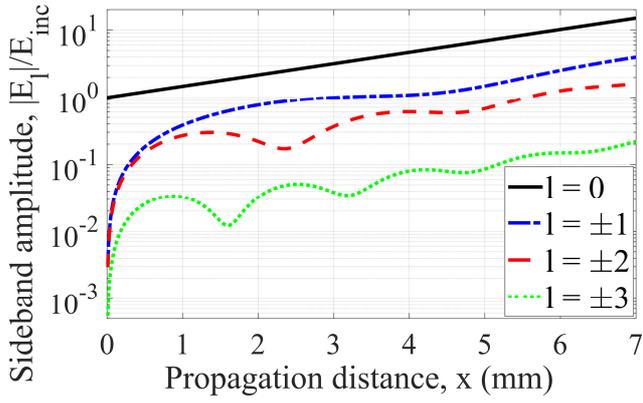

Fig. 1. (Color online). The amplitudes of spectral components of the X-ray field according to the analytical solution Eq. (21) vs. the propagation distance in the active medium of $C^{5+}$ ions, $x$, for $\tau = 208$ fs (which corresponds to the maximum amplification, see Fig. 2) and $P_\Omega = 4.1$. Black solid line shows the central spectral component, $l=0$. Blue dash-dot, red dashed, and green dotted curves correspond to the sidebands with $l=\pm 1$, $l=\pm 2$, and $l=\pm 3$, respectively.

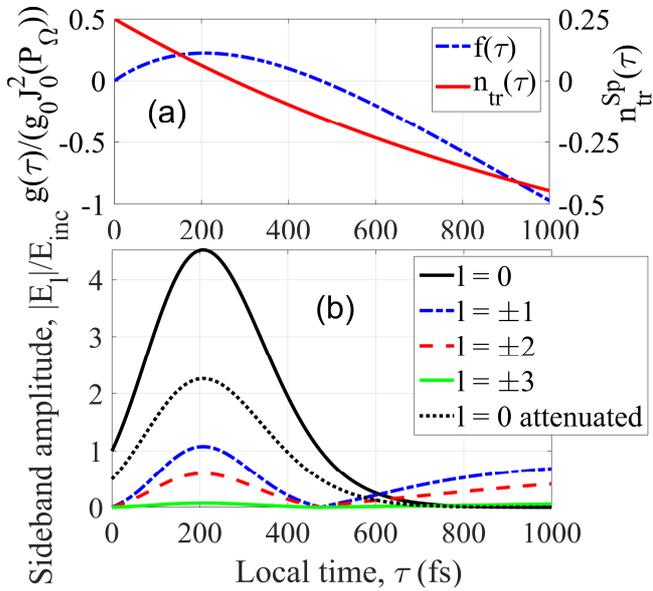

Fig. 2. (Color online). (a) Local-time dependencies of (i) the population difference at the lasing transitions, $n_{tr}^{Sp}(\tau)$, see Eq. (16), solid red curve, right vertical axis, and (ii) the normalized effective gain coefficient, $g(P_\Omega, \tau)/\left[g_0 J_0^2(P_\Omega)\right]$, see Eq. (21c), dash-dot blue curve, left vertical axis. (b) Local-time dependencies of the amplitudes of the central spectral component, $l=0$, solid black curve, and the sidebands of the X-ray field: $l=\pm 1$, shown by blue dash-dot curve, $l=\pm 2$, shown by red dashed curve, and $l=\pm 3$, shown by green solid curve. Black dotted curve shows the central spectral component, $l=0$, after a two-time attenuation. The amplitudes of spectral components are calculated via the analytical solution (21) at the output from the active plasma medium of the length $L = 3.9$ mm. The modulation index is $P_\Omega = 4.1$.

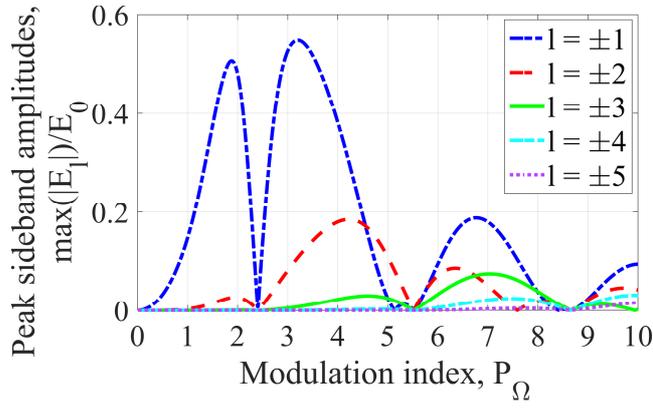

Fig. 3. (Color online). The maximum achievable values of the amplitudes of sidebands relative to the amplitude of the resonant (zeros) spectral component of the X-ray field, calculated via the analytical solution (21) at $\tau = \tau_{max} \approx 208$ fs, as a function of the modulation index. Blue dash-dot curve corresponds to $l=\pm 1$, red dashed curve corresponds to $l=\pm 2$, green solid curve shows $l=\pm 3$, cyan dash-dot and magenta dotted curves represents the cases of $l=\pm 4$ and $l=\pm 5$, respectively. Generally, the peak amplitudes of sidebands are achieved at different propagation distances for each sideband number and each value of the modulation index.

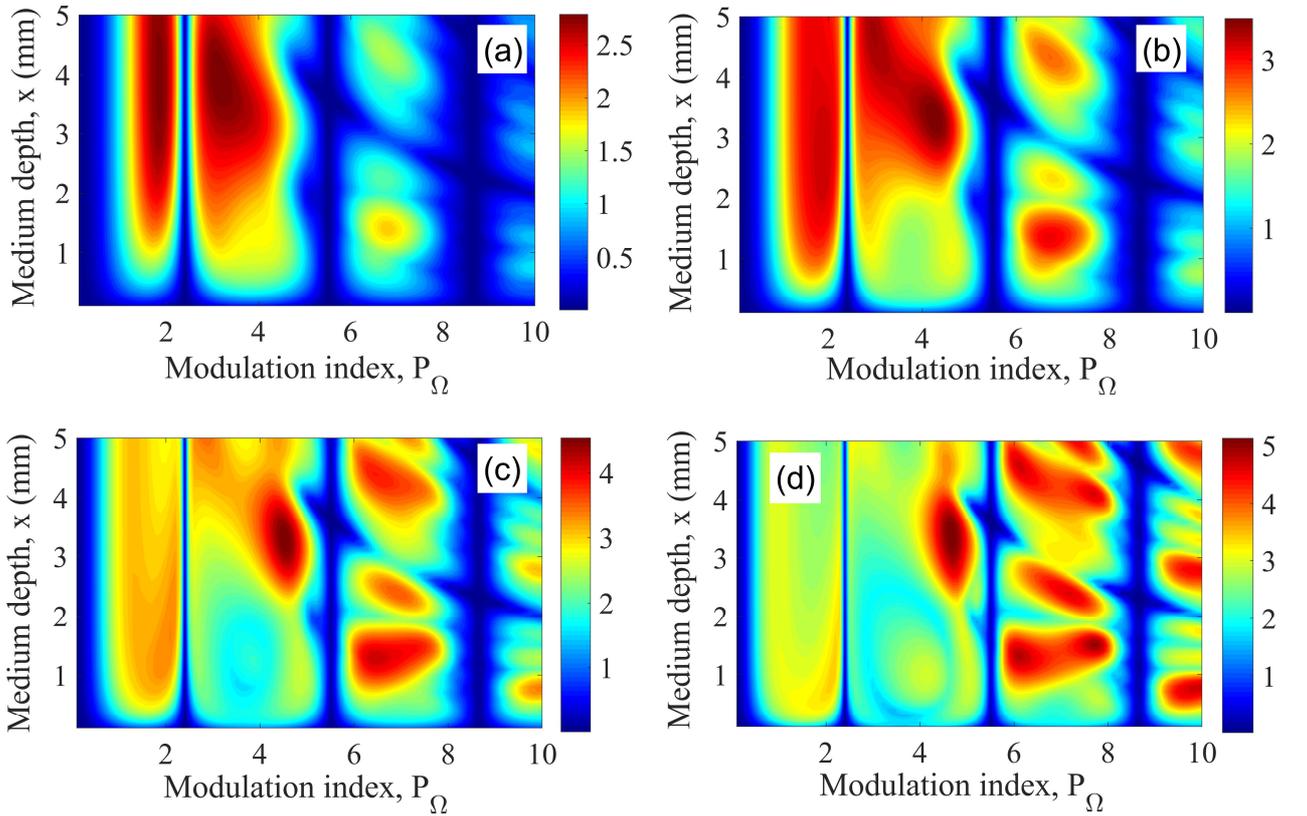

Fig. 4. (Color online). (a) The dependence of contrast of the X-ray field on the propagation distance through the medium, $x$, and on the modulation index, $P_\Omega$. (b), (c) and (d) The same as in (a), but the central spectral component of the output X-ray field is externally attenuated by 2 times ($K=2$) in (b), 4 times ($K=4$) in (c) and 8 times ($K=8$) in (d). Note that the color scale in different panels is different. The maximum achievable values of contrast are: $C \approx 2.8$ at $x=3.9$ mm and $P_\Omega = 3.2$ in (a); $C \approx 3.5$ at $x=3.3$ mm and $P_\Omega = 4.3$ in (b); $C \approx 4.6$ at $x=3.3$ mm and $P_\Omega = 4.5$ in (c); and $C \approx 5.2$ at $x=3.4$ mm and $P_\Omega = 4.7$ in (d).

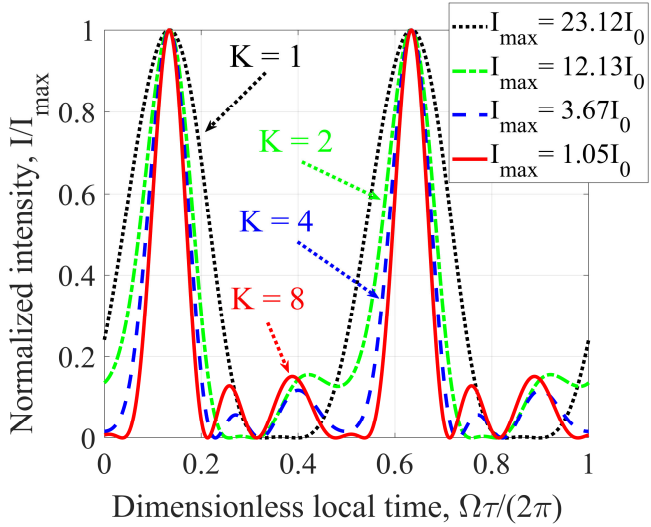

Fig. 5. (Color online). The time-dependence of intensity of the output X-ray field near the peak of its envelope at $\tau = 208$ fs, calculated via the analytical solution (21). Black dotted curve corresponds to the X-ray field as it is (no spectral filtering), $K$=1. Green dash-dot curve: the central spectral component is externally attenuated by two times, $K$=2. Blue dashed curve: the central spectral component is attenuated four times, $K$=4. Red solid curve: the resonant spectral component is attenuated eight times, $K$=8. The intensities are normalized on their maximum values, which are shown at the legend. In each case the length of the medium and the index of modulation are chosen to be optimal for the corresponding level of attenuation, that is $x$=3.9 mm and $P_\Omega = 3.2$ for $K$=1; $x$=3.3 mm and $P_\Omega = 4.3$ for $K$=2; $x$=3.3 mm and $P_\Omega = 4.5$ for $K$=4; $x$=3.4 mm and $P_\Omega = 4.7$ for $K$=8 (see Fig. 4).

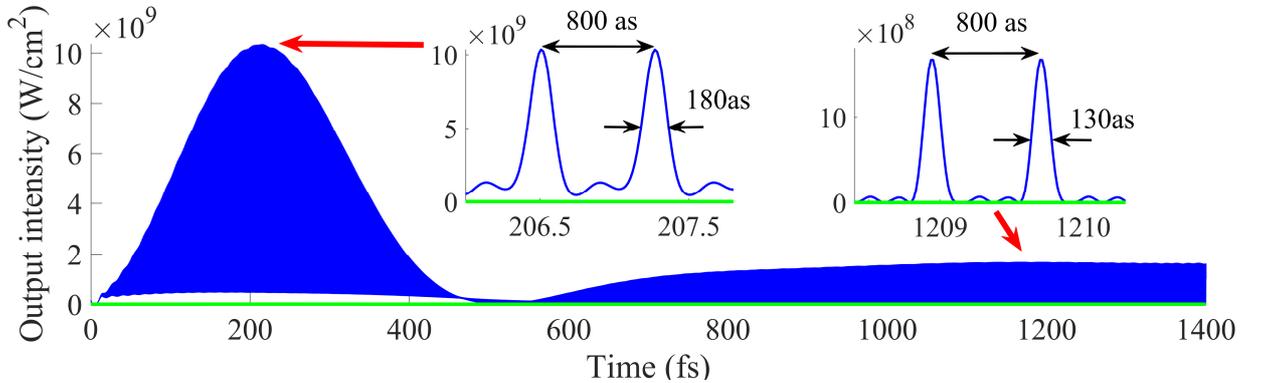

Fig. 6. (Color online). The numerical solution for the time-dependence of intensity of the X-ray field at the output from the medium of the length $x$=3.9 mm; the modulation index is $P_\Omega = 4.1$. The central spectral component is externally attenuated two times. The inserts show the shape of pulses at the peak of the intensity envelope and at the tail of the pulse train. In the first case ($\tau \approx$ 208 fs) the pulses are produced in the amplifying medium, while in the second case ($\tau \approx$ 1210 fs) the medium is absorbing for the X-ray field.

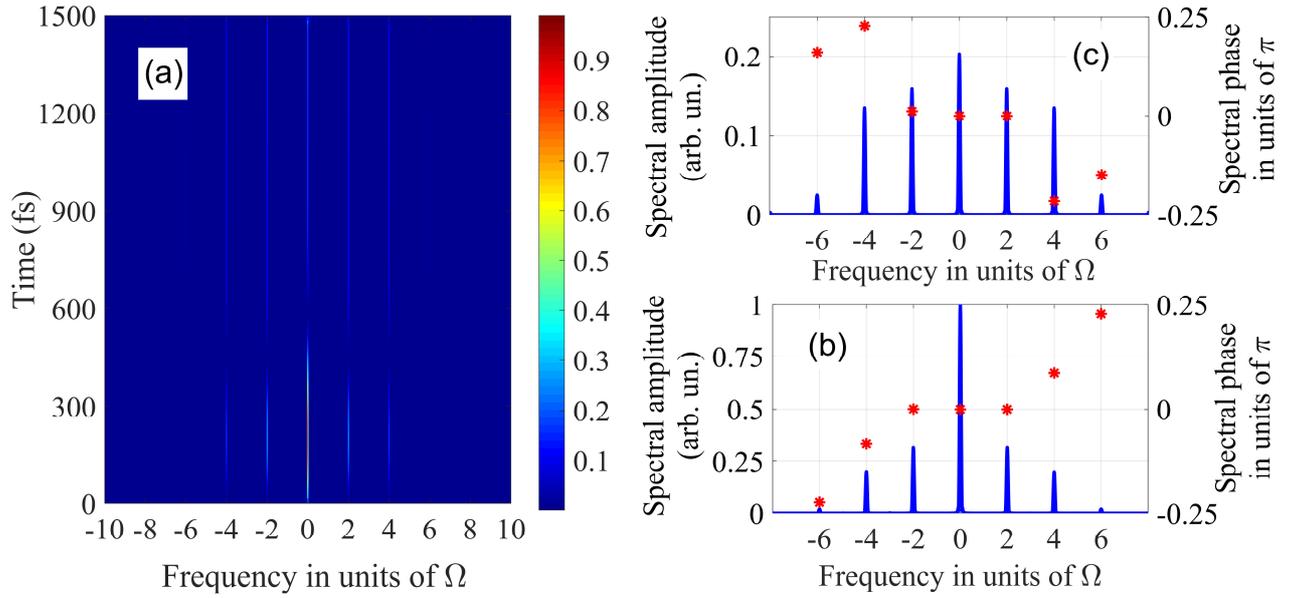

Fig. 7. (Color online). (a) The time-windowed Fourier transform (an amplitude spectrum) of the X-ray field shown in Fig. 6 (a numerical solution for $x$=3.9 mm, $P_\Omega = 4.1$, the resonant spectral component is attenuated two times). (b) Blue curve, left vertical axis – a cut of Fig. 7(a) for $\tau$ = 215 fs (the time-windowed spectrum, which corresponds to the left insert in Fig. 6). Red stars, right vertical axis – the central phases of the spectral components. (c) The same as in (b) but for $\tau$ = 1210 fs (the time-windowed spectrum, which corresponds to the right insert in Fig. 6). The spectral amplitude in (a) is normalized to its maximum value, which is achieved at $\tau$ = 215 fs. In (b) and (c) the normalization factor is the same as in (a).

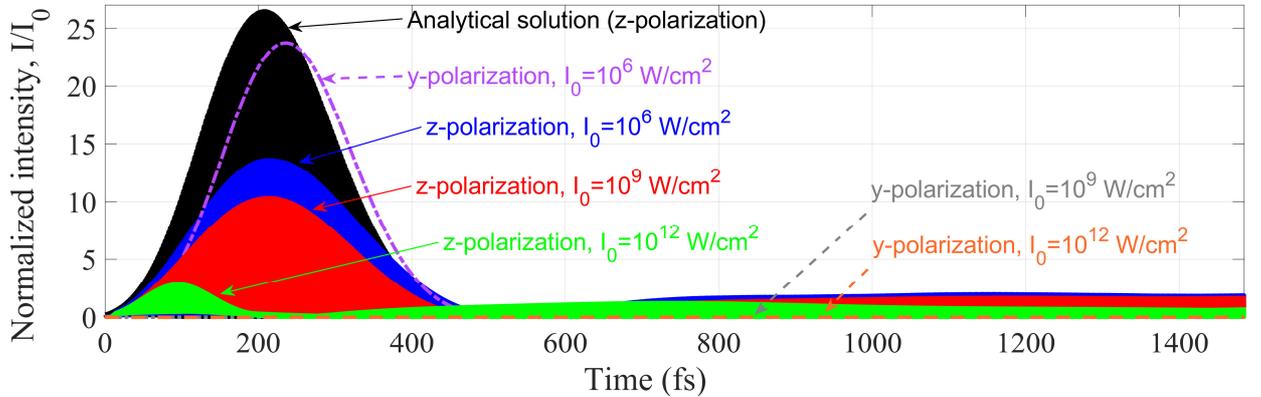

Fig. 8. (Color online). The time-dependencies of intensity of the polarization components of the X-ray field, calculated analytically and numerically for different intensities $I_0$ of the z-polarized seeding field. The parameters of the medium are the same as in Fig. 6: the length of the medium is $x$=3.9 mm; the modulation index is $P_\Omega = 4.1$. The central spectral component of the field is attenuated two times. Black solid curve shows the analytical solution for z-polarization component of the amplified X-ray field. Blue, red and green solid curves show the results of numerical calculations for z-polarization component of the field, assuming $I_0 = 10^6$ W/cm$^2$, $I_0 = 10^9$ W/cm$^2$, and $I_0 = 10^{12}$ W/cm$^2$, respectively. The results of numerical calculations for y-polarization component of the X-ray field for $I_0 = 10^6$ W/cm$^2$, $I_0 = 10^9$ W/cm$^2$, and $I_0 = 10^{12}$ W/cm$^2$, are shown by dash-dot lavender curve, dotted grey curve and orange dashed curve (the latter two are almost indistinguishable). The output intensity is normalized to the intensity of the seeding field, $I_0$.

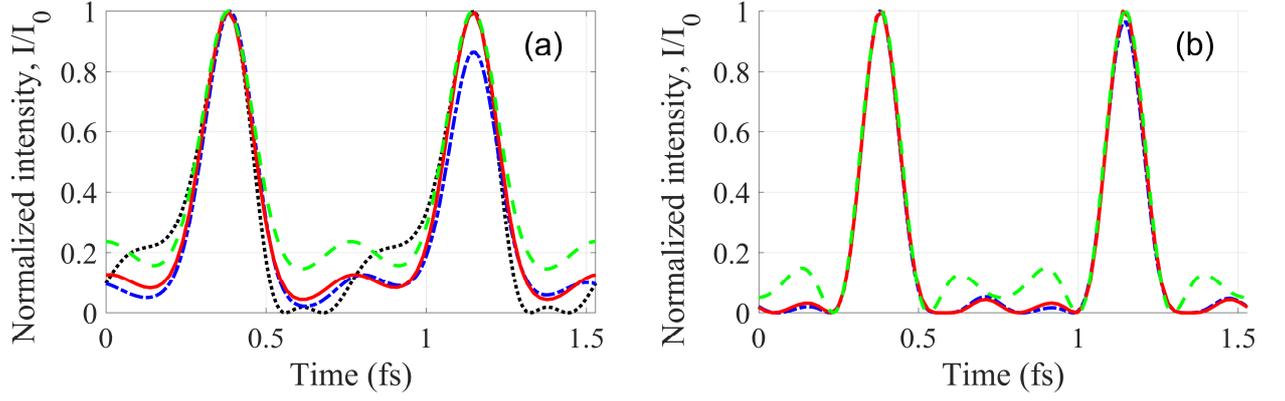

Fig. 9. (Color online). The shape of individual pulses, produced under the conditions of Fig. 8. Black dotted curve is the analytical solution. Blue dash-dot, red solid and green dashed curves are the results of numerical calculations for the intensities of the seeding field $I_0 = 10^6$ W/cm$^2$, $I_0 = 10^9$ W/cm$^2$, and $I_0 = 10^{12}$ W/cm$^2$, respectively. Panels (a) and (b) show the time-dependencies of intensity of the X-ray field at the peak of the intensity envelope (a), and at the second local maximum of the intensity envelope (b); compare with inserts in Fig. 6. The position of the maxima differs for different solutions. The analytical solution in (a) is plotted for $\tau \approx 208$ fs. The numerical solutions in (a) and (b) are plotted for $\tau \approx 215$ fs and $\tau \approx 1160$ fs (for $I_0 = 10^6$ W/cm$^2$), $\tau \approx 210$ fs and $\tau \approx 1180$ fs (for $I_0 = 10^9$ W/cm$^2$), as well as $\tau \approx 95$ fs and $\tau \approx 740$ fs (for $I_0 = 10^{12}$ W/cm$^2$). In each case the intensity is normalized on its peak value.